\documentclass[doublecol]{epl2}
\usepackage{amssymb,xcolor,graphicx,amsmath,booktabs,citesort,float,multirow}

\title{Group-based ranking method for online rating systems with spamming attacks}
\shorttitle{A group-based ranking method for online rating systems}
\author{Jian Gao\inst{1} \and Yu-Wei Dong\inst{1} \and Ming-Sheng Shang\inst{1}\footnote{E-mail: msshang@uestc.edu.cn} \and Shi-Min Cai\inst{1} \and Tao Zhou\inst{1,2}\footnote{E-mail: zhutou@ustc.edu}}
\shortauthor{J. Gao \etal}

\institute{
    \inst{1} Web Sciences Center, University of Electronic Science and Technology of China - 611731 Chengdu, PRC\\
    \inst{2} Big Data Research Center, University of Electronic Science and Technology of China - 611731 Chengdu, PRC}


\abstract{Ranking problem has attracted much attention in real systems. How to design a robust ranking method is especially significant for online rating systems under the threat of spamming attacks. By building reputation systems for users, many well-performed ranking methods have been applied to address this issue. In this Letter, we propose a group-based ranking method that evaluates users' reputations based on their grouping behaviors. More specifically, users are assigned with high reputation scores if they always fall into large rating groups. Results on three real data sets indicate that the present method is more accurate and robust than correlation-based method in the presence of spamming attacks.}

\begin{document}
\maketitle

\section{Introduction}

With the rapid development of the Internet, billions of services and objects are online for us to choose \cite{Lu2012}. At the same time, the problem of information overload troubles us every day \cite{Zeng2012a,Zeng2014,Guo2014}. Therefore, many web sites (Amazon, Ebay, MovieLens, Netlfix, etc.) introduce online rating systems, where users can give discrete ratings to objects. In turn, the ratings of an object serve as a reference and latter affect other users' decisions \cite{Jindal2007,Muchnik2013}. Basically, high ratings can promote the consumption, while low ratings play the opposite role \cite{Yu2012}. In real cases, some users may give unreasonable ratings since they are simply unfamiliar with the related field \cite{Pan2013}, and some others deliberately give biased ratings for various psychosocial reasons \cite{Wang2011,Chung2013,Yang2012,Huang2012,Zhang2015}. These widely existed distort ratings can harm or boost objects' reputation, mislead others' judgments, and affect the evolution of rating systems \cite{Zeng2012b,Zhang2013,Zhao2014}. Due to the negative effects of spamming attacks, how to design a robust method for online rating systems is becoming an urgent task \cite{Mukherjee2011,Fei2013,Lin2014}.

To solve this problem, normally, building a reputation system for users is a good way \cite{Resnick2000,Masum2004,Josang2007,Allahbakhsh2013,Ling2013,Laureti2006,de2007,Zhou2011,Liao2014}. Laureti \etal~\cite{Laureti2006} proposed an iterative refinement (IR) method, where a user's reputation is inversely proportional to the difference between his ratings and the estimation of the corresponding objects' quality (\emph{i.e.}, weighted average rating). The reputation and the estimated quality are iteratively calculated until they become stable. An improved IR method is proposed in \cite{de2007}, by assigning trust to each individual rating. Later, Zhou \etal~\cite{Zhou2011} proposed the correlation-based ranking (CR) method that is robust to spamming attacks, where a user's reputation is iteratively determined by the correlation between his ratings and objects' estimated quality. Very recently, by introducing a reputation redistribution process and two penalty factors, Liao \etal~\cite{Liao2014} further improved the CR method.

\begin{figure*}
\centering
\onefigure[width=150mm]{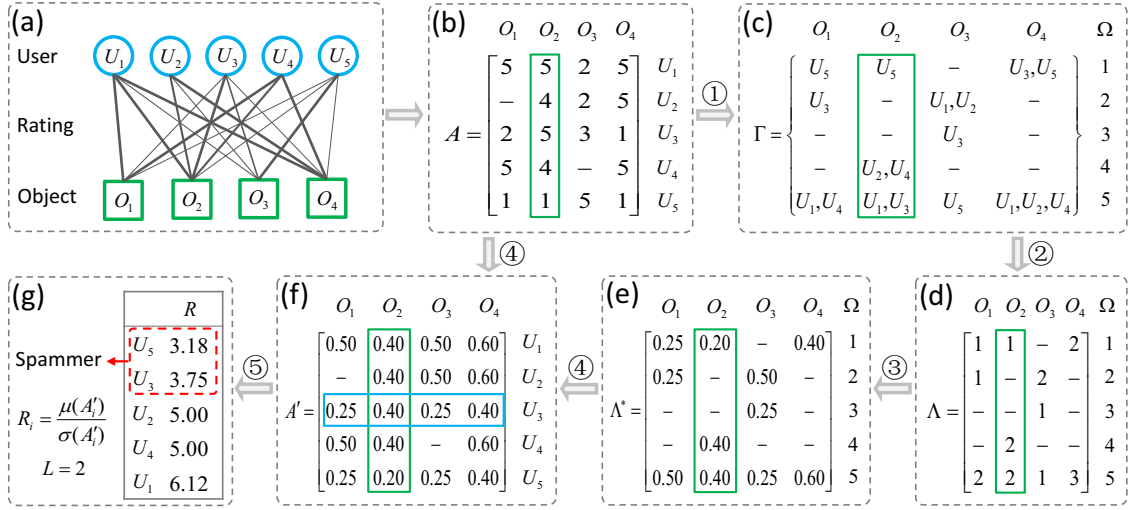}
    \caption{(Color online) Illustrating the group-based method. The number besides the gray arrow marks the step of the procedure. (a) The original weighed bipartite network, $G$. (b) The corresponding rating matrix, $A$. The row and column correspond to users and objects, respectively. The symbol ``-" stands for a non-rating, which can be ignored in the calculation. (c) The groups of users, $\Gamma$, after being grouped according to their ratings. Take $O_{2}$ as an example (green vertical box). As $U_{2}$ and $U_{4}$ rated 4 to $O_{2}$, they are put into group $\Gamma_{4,2}$. (d) The sizes of groups, $\Lambda$, \emph{e.g.} $\Lambda_{4,2}=2$ as $\Gamma_{4,2} = \{ U_{2}, U_{4}\}$. (e) The rating-rewarding matrix, $\Lambda^{\ast}$, constructed by normalizing $\Lambda$ by column, \emph{e.g,} $\Lambda^{\ast}_{4,2}=2/(1+2+2)=0.40$. (f) The rewarding matrix, $A'$, obtained by mapping matrix $A$ referring to $\Lambda^{\ast}$, \emph{e.g.} $A'_{4,2}=0.40$. (g) The ranking list of users based on reputation $R$. Take $U_{3}$ as an example (blue horizontal box in (f)), $R_{3}=\mu(A'_{3})/\sigma(A'_{3})=3.75$. Setting spam list's length as $L=2$, then $U_{5}$ and $U_{3}$ (red dashed box) are detected as spammers.}
\label{FIG:1GR}
\end{figure*}

In the majority of previous works \cite{Raykar2010,Tian2012}, a single standard objects' quality is required in determining users' reputations, with an underlying assumption that every object is associated with a most objective rating that best reflect its quality. However, in real cases, one object may have multiple valid ratings, since the ratings are subjective and can be affected by users' background \cite{Shi2009,Welinder2010,Reed2010,Ross2010}. In the presence of more than one reasonable answer to a single task, Tian \etal \cite{Tian2012} analyzed the group structure of schools of thought in solving the problem of identifying reliable workers as well as unambiguous tasks in data collection. Specifically, a worker who is consistent with many other workers in most of the tasks is reliable, and a task whose answers form a few tight clusters is easy and clear. Analogously, in the online rating systems, one object's quality is clear if its ratings are centralized, while it is not clear if the ratings are widely distributed. Under this framework, a single estimation of an object's quality is no longer applicable \cite{Raykar2010}. Practically, a random rating to objects with confusing quality should be acceptable since no single rating can dominant its true quality, while a biased rating to objects with clear quality is unreasonable. Users who are consistent with the majority in most ratings will form big groups and be trusted since herding is a well-documented feature of human behaviour \cite{Muchnik2013,Raafat2009}. Users who always give distort ratings will form relatively small groups and be highly suspected since unreasonable or biased ratings are discordant \cite{Tian2012}. These ideas bring us a promising way to build reputation systems based on users' grouping behaviour instead of solving the crucial problem of estimating objects' true qualities as before.

In this Letter, we propose a group-based ranking (GR) method for online rating systems with spamming attacks. By grouping users according to their ratings, users' reputations are determined according to the corresponding group sizes. If they always fall into large groups, their reputations are high, on the contrary, their reputations are low. Extensive experiments on three real data sets (MovieLens, Netflix and Amazon) suggest that the proposed method outperforms the CR method.

\section{Method}
\label{Sec:model}

The online rating system is naturally described by a weighed bipartite network $G=\{U, O, E\}$, where $U=\{U_{1}, U_{2}, ..., U_{m}\}$, $O=\{O_{1}, O_{2}, ..., O_{n}\}$, $E=\{E_{1}, E_{2}, ..., E_{l}\}$ are sets of users, objects and ratings, respectively \cite{Shang2010}. The degree of a user $i$ and an object $\alpha$ are denoted as $k_i$ and $k_{\alpha}$, respectively. Here, we use Greek and Latin letters, respectively, for object-related and user-related indices to distinguish them. Considering a discrete rating system, the bipartite network can be represented by a rating matrix $A$ \cite{Zhou2007b}, where the element $a_{i\alpha} \in \Omega=\{\omega_{1}, \omega_{2}, ..., \omega_{z}\}$ is the weight of the link connecting node $U_{i}$ and node $O_{\alpha}$, \emph{i.e.}, the rating given by user $i$ to object $\alpha$. In a reputation system, each user $i$ will be assigned a reputation, denoted as $R_{i}$. The users with very low reputations are detected as spammers.

The GR method works as follows. Firstly, we group users according to their ratings. Specifically, for an object $\alpha$, we put users who gave the rating $\omega_{s}$ into group $\Gamma_{s\alpha}$:
\begin{equation}
\label{eq:G}
    \Gamma_{s\alpha}=\{U_{i}\mid a_{i\alpha}=\omega_{s}, i=1,2,...,m\}.
\end{equation}
Obviously, user $i$ belongs to $k_{i}$ different groups. Secondly, we calculate the sizes of all groups $\Lambda_{s\alpha}=|\Gamma_{s\alpha}|$, \emph{i.e.} the number of users who gave the rating $\omega_{s}$ to object $\alpha$. Thirdly, we establish a matrix $\Lambda^{\ast}$, named rating-rewarding matrix, by normalizing $\Lambda$ per column:
\begin{equation}
\label{eq:L}
    \Lambda^{\ast}_{s\alpha}=\frac{\Lambda_{s\alpha}}{k_{\alpha}}.
\end{equation}
Fourthly, referring to $\Lambda^{\ast}$, we map the original rating matrix $A$ to a matrix $A'$, named rewarding matrix. More specifically, the rewarding that a user $i$ obtain from his rating $a_{i\alpha}$ is defined as $A'_{i\alpha}=\Lambda^{\ast}_{s\alpha}$, where $a_{i\alpha}=\omega_{s}$. $A'_{i\alpha}$ is null if user $i$ has not yet rated object $\alpha$.

Then, we assign reputations to users according to their rewarding. On the one hand, if the average of a user's rewarding is small, most of his ratings must be deviated from the majority, suggesting that he is highly suspected. On the other hand, if the rewarding varies largely, he is also untrustworthy for his unstable rating behavior. Based on these considerations, we defined user $i$'s reputation as
\begin{equation}
\label{eq:R}
    R_{i}=\frac{\mu(A'_{i})}{\sigma(A'_{i})},
\end{equation}
where $\mu$ and $\sigma$ are functions of mean value and standard deviation, respectively. Specifically, the mean value of $A'_{i}$ is defined as
\begin{equation}
\label{eq:mu}
    \mu(A'_{i})=\sum_{\alpha}\frac{A'_{i\alpha}}{k_{i}},
\end{equation}
and the standard deviation of $A'_{i}$ is defined as
\begin{equation}
\label{eq:sigma}
    \sigma(A'_{i})=\sqrt{\frac{\sum_{\alpha}(A'_{i\alpha}-\mu(A'_{i}))^2}{k_{i}}}.
\end{equation}
In fact, $R_{i}$ is the same with the inverse of the coefficient of variation \cite{Lin1989} of vector $A'_{i}$, which shows the dispersion of the frequency distribution of user $i$'s rewardings. Finally, we sort users in ascending order by reputation, and deem the top-$L$ ones as detected spammers. A visual representation of GR method is given in fig.~\ref{FIG:1GR}.

\begin{table}
\centering
    \caption{Basic statistics of the three real data sets. $m$ is the number of users, $n$ is the number of objects, $\langle k_{U}\rangle$ is the average degree of users, $\langle k_{O}\rangle$ is the average degree of objects, and $S=l/mn$ is the sparsity of the bipartite network.}
    \begin{tabular}{llllll}
    \toprule
    Data set & $m$ & $n$ & $\langle k_{U}\rangle$ & $\langle k_{O}\rangle$ & $S$ \\
    \midrule
    MovieLens & 943 & 1682 & 106 & 60 & 0.063\\
    Netflix & 1038 & 1215 & 47 & 40 & 0.039\\
    Amazon & 662 & 1500 & 36 & 15 & 0.023\\
    \bottomrule
    \end{tabular}
\label{Table:1dataset}
\end{table}

\section{Data and Metrics}
\label{Sec:DM}
We consider three commonly studied real data sets, MovieLens, Netflix and Amazon, to test the accuracy of GR method. MovieLens and Netflix contain ratings on movies, provided by GroupLens project at University of Minnesota (www.grouplens.org) and released by the DVD rental company Netflix for a contest on recommender systems (www.netflixprize.com), respectively. Amazon contains ratings on products (\emph{e.g.} books, music, etc) crawled from amazon.com \cite{Jindal2008}. All the three data sets use a 5-point rating scale with 1 being the worst and 5 being the best. Herein, we sampled and extracted three smaller data sets from the original data sets, respectively, by choosing users who have at least 20 ratings and objects having been rated by these users since it's hard to tell whether small-degree users are spammers \cite{Liao2014}. The basic statistics of data sets are summarized in table~\ref{Table:1dataset}.

\subsection{Generating artificial spammer}
Two types of distorted ratings are widely found in real rating systems, namely, malicious ratings and random ratings. The malicious ratings are from spammers who always gives minimum (maximum) allowable ratings to push down (up) certain target objects \cite{Wang2011,Chung2013}. The random ratings mainly come from some naughty users or test engineers who randomly give meaningless ratings \cite{Zeng2012b,Zhang2013}. As spammers are unknown in real data, to test the method, we manipulate the three real data sets by adding either type of artificial spammers (\emph{i.e.} malicious or random) at one time.

In the implementation, we randomly select $d$ users and assign them distorted ratings: (i) integer 1 or 5 with the same probability (\emph{i.e.}, 0.5) for malicious spammers, and (ii) random integers in $\{ 1,2,3,4,5 \}$ for random spammers. Thus, the ratio of spammers is $q=d/m$. To study the effects of spammers' activity, we define $p = k/n$ the activity of spammers, where $k$ is the degree of each spammer. Here $k$ is a tunable parameter, that is, if $k$ is no more than a spammer's original degree, we randomly select his/her $k$ ratings and replace them with distorted ratings and the un-selected ratings are ignored. Otherwise, after replacing all the spammer's original ratings, we randomly select remaining number of non-rated objects and assign them distorted ratings.

\begin{figure}
\centering
\onefigure[width=75mm]{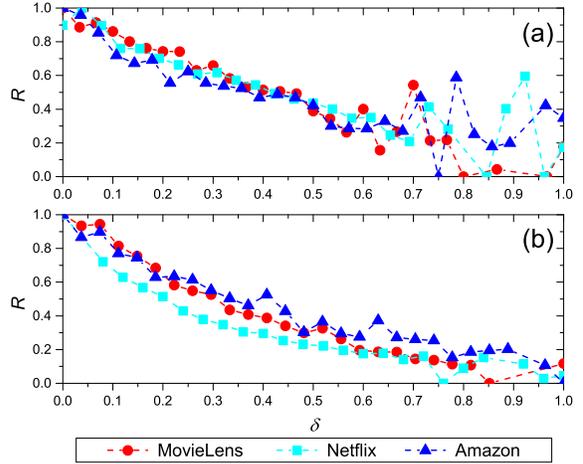}
    \caption{(Color online) The relation between reputation $R$ and rating error $\delta$ (bins) in different methods. (a) and (b) are for the CR and GR method, respectively. $\delta$ and $R$ are respectively normalized for comparison under different data sets.}
\label{FIG:2rou}
\end{figure}

\subsection{Metrics for evaluation}
To evaluate the performance of ranking methods, we adopt two commonly used metrics: recall \cite{Herlocker2004} and AUC (the area under the ROC curve) \cite{Hanley1982}. The recall value measures to what extent the spammers can be detected in the top-$L$ ranking list,
\begin{equation}
\label{eq:P}
    R_{c}(L)=\frac{d'(L)}{d},
\end{equation}
where $d'(L)\leq d$ is the number of detected spammers in the top-$L$ list. A higher $R_{c}$ indicates a higher accuracy.

Note that, $R_{c}$ only focuses on the top-$L$ ranks, and thus we also consider an $L$-independent metric called AUC. Provided the rank of all users, AUC value can be interpreted as the probability that a randomly chosen spammer is ranked higher than a randomly chosen non-spammer. To calculate AUC, at each time we randomly pick a spammer and a non-spammer to compare their reputations, if among $N$ independent comparisons, there are $N'$ times the spammer has a lower reputation and $N''$ times they have the same reputation, the AUC value is
\begin{equation}
\label{eq:AUC}
    AUC=\frac{N' + 0.5N''}{N}.
\end{equation}
If all users are ranked randomly, the AUC value should be about 0.5. Therefore, the degree to which the value exceeds 0.5 indicates how better the method performs than pure chance \cite{lu2011}.

\begin{figure}
\centering
\onefigure[width=80mm]{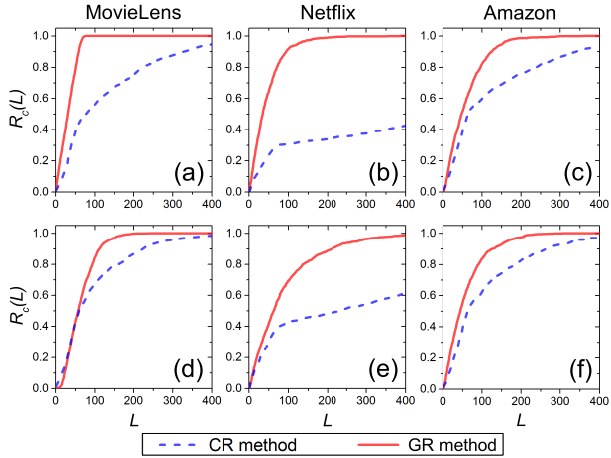}
    \caption{(Color online) The recall $R_{c}$ of different algorithms as a function of the length of the list $L$. (a), (b) and (c) are for malicious spammers, (d), (e) and (f) are for random spammers with $d=50$ being fixed. The parameter $p$ is set as about 0.05, 0.03 and 0.02 for MovieLens, Netflix and Amazon, respectively. The results are averaged over 100 independent realizations.}
\label{FIG:3PL}
\end{figure}

\section{Results}
\label{Sec:Res}

According to the single standard assumption, each object $\alpha$ has a true quality, denoted by $Q_{\alpha}$. As the true quality is unknown in reality, taking the average rating as an estimation of $Q_{\alpha}$ is the most straightforward way. Then, the rating error of user $i$ is defined as
\begin{equation}
\label{eq:DE}
    \delta_{i}=\frac{\sum_{\alpha}|a_{i\alpha}-\hat{Q}_{\alpha}|}{k_{i}},
\end{equation}
where $\hat{Q}_{\alpha}=1/\sum_{j \in \Gamma_{\alpha}}a_{j\alpha}$ is the average rating of object $\alpha$ and $\alpha$ runs over all objects being rated by user $i$. A reasonable ranking method should give high reputations to the users with low rating errors, \emph{i.e.} $R_{i}$ should be negatively correlated with $\delta_{i}$.

Figure~\ref{FIG:2rou} shows the relation between users' rating errors and reputations evaluated by CR and GR, respectively. As users' rating errors are continuous and with different scales, we normalize and divide them into bins with the length 0.05. As one can see, the two methods both assign high reputations for users with small $\delta_{i}$, while GR method outperforms CR method by stably assigning low reputations (see fig.~\ref{FIG:2rou}(b)) for users with large $\delta_{i}$. Furthermore, to quantify the correlation, we calculate the Pearson correlation coefficient $\rho$ \cite{Lin1989} between $R_{i}$ and $\delta_{i}$. Specifically, $\rho=-0.956$ ($-0.949$), $-0.906$ ($-0.872$) and $-0.966$ ($-0.816$) after applying GR (CR) method to MovieLens, Netflix and Amazon, respectively. The larger negative correlation suggest that GR method is better on evaluating users' reputations.

\begin{table}
\centering
    \caption{AUC of different algorithms for the real data sets with $d=50$ being fixed. The parameter $p$ is set as about 0.05, 0.03 and 0.02 for MovieLens, Netflix and Amazon, respectively. The results are averaged over 100 independent realizations.}
    \begin{tabular}{lcccc}
    \toprule
    \multirow{2}{*}{Data set} & \multicolumn{2}{c}{Malicious spamers} & \multicolumn{2}{c}{Random spamers}\\
    \cline{2-5}
      & CR & GR & CR & GR\\
    \midrule
    MovieLens & 0.876 & 0.994 & 0.914 & 0.959\\
    Netflix & 0.543 & 0.977 & 0.668 & 0.930\\
    Amazon & 0.824 & 0.941 & 0.877 & 0.949\\
    \bottomrule
    \end{tabular}
\label{Table:2eff}
\end{table}

\subsection{Effectiveness and efficiency}
To test the effectiveness of ranking algorithms, based on the three real data sets, we first generate artificial data sets with 50 spammers (\emph{i.e.} $d=50$). Each data set is only with one type of spammers: malicious or random. On the generated data sets, we calculate recall of different algorithms as a function of the spammer list's length $L$. As shown in fig.~\ref{FIG:3PL}, GR method has remarkable advantage over the CR method on detecting both types of spammers, especially when $L$ is larger than $d$. We also note that $R_{c}$ of ranking malicious spammers is a little higher than that of random spammers when $L$ is smaller than $d$, which implies that to detect random spammers is relatively harder.

Results of AUC values are shown in table~\ref{Table:2eff}, where one can see that AUC values of GR method is higher than that of the CR method for every data set, suggesting that GR method has significant advantage towards the CR method. It also shows that the CR method is better at detecting random spammers than malicious spammers, while GR method is inverse. In addition, it is worthy to be noticed that the AUC is generally lower in Netflix, especially for the CR method. One possible explanation is that there are more harmful spammers in Netflix and the CR method is very sensitive to ``real" spammers \cite{Zhou2011,Liao2014}. Additionally, in Netflix, there are more small degree objects whose quality will be considered higher by the CR method \cite{Liao2014}, which may also lead to the biased ranking.

\subsection{Robustness against spammers}
\begin{figure}
\centering
\onefigure[width=85mm]{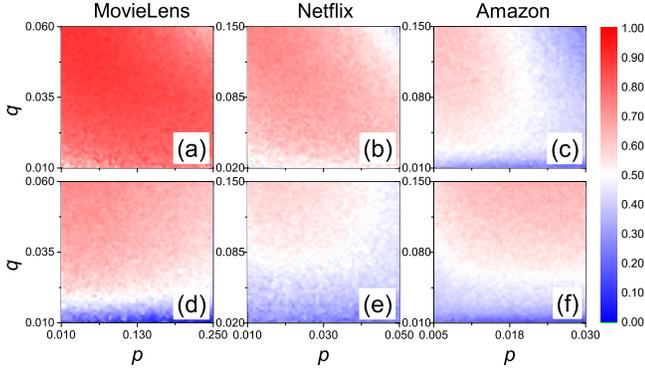}
    \caption{(Color online) The effectiveness of GR method. The color marks recall $R_{c}$. $q$ and $p$ are ratio of spammers and ratio of objects rated by spammers, respectively. (a), (b) and (c) are for malicious spammers. (d), (e) and (f) are for random spammers. The parameter is set as $L=d$.  The results are averaged over 100 independent realizations.}
\label{FIG:4P}
\end{figure}

We then study the robustness of different methods by varying $p$ (the ratio of objects rated by spammers) and $q$ (the ratio of spammers). In the following, we set the length of detected spam list being equal to the number of artificial spammers, namely, $L=d$. Figure~\ref{FIG:4P} shows the recall obtained by GR method. The ranges of $p$ and $q$ are personalized set for different data sets referring to their sparsity. One can observe that, overall, GR method has better performance on detecting malicious spammers, especially when $p$ is small (\emph{i.e.} spammers are of small degree). Moreover, when $q$ is small, the recalls of ranking random spammers are low for MovieLens and Netflix data sets (see figs.~\ref{FIG:4P}(d) and \ref{FIG:4P}(e)), while for Amazon data set, the recall of ranking both two types of spammers is low (see figs.~\ref{FIG:4P}(c) and \ref{FIG:4P}(f)). In addition, the recall positively increases with $q$. These observations suggest that (i) detecting malicious spammers is easier than random spammers, (ii) MovieLens and Netflix may contain more ``real" spammers, and (iii) GR method is powerful to detect spammers who only rate a small number of objects.

\begin{figure}
\centering
\onefigure[width=85mm]{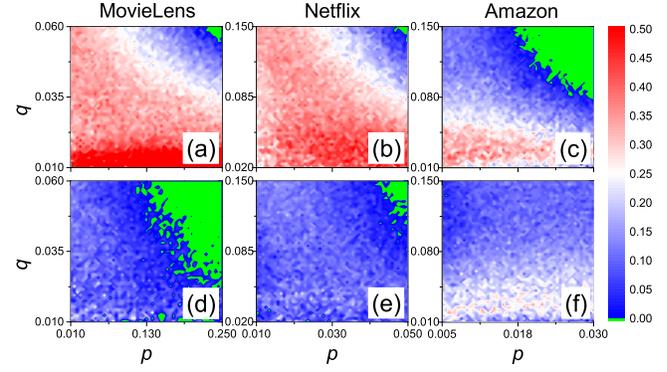}
    \caption{(Color online) The comparison of GR and CR methods. The color marks $\Delta R_{c}$ if $\Delta R_{c}>0$, otherwise the color is green, meaning that the CR method is better. (a), (b) and (c) are for malicious spammers. (d), (e) and (f) are for random spammers. The parameter is set as $L=d$.  The results are averaged over 100 independent realizations.}
\label{FIG:5DifP}
\end{figure}

To comprehensively compare the performance of GR and CR methods, we calculate the difference of recall between the two, formulateed as $\Delta R_{c}=R_{c}^{GR}-R_{c}^{CR}$. As shown in fig.~\ref{FIG:5DifP}, $\Delta R_{c}$ is above 0 in most area, suggesting that the overall performance of GR method is better. In detail, GR method has remarkable advantage in detecting malicious spammers, while the advantage is not obvious for random spammers. In addition, $\Delta R_{c}$ is big when $p$ and $q$ are small, implying that GR method is more robust against a small number of small-degree spammers, which are usually difficult to be detected out.

\section{Conclusions and Discussion}
\label{CD}

In summary, we have proposed a group-based (GR) method to solve the ranking problem in online rating systems with spamming attacks. By grouping users according to their ratings, we construct a rating-rewarding matrix according to the corresponding group sizes, base on which we map a user's rating vector to rewarding vector. Then, this user's reputation is assigned according to the inverse of dispersion of frequency distribution of his rewarding vector.  Extensive experiments showed that the proposed method is effective on evaluating users' reputations, especially for those with high rating errors. In testing with the generated data with two types of artificial spammers, GR method gives higher performance in both accuracy and robustness compared with the correlation-based ranking (CR) method, especially on resisting small-degree spammers. Interestingly, the accuracy of spam detection on Netflix data set is low for both the two methods, indicating that there are more original distort ratings in Netflix data sets, which is in accordance with some previous studies \cite{Zhou2011,Liao2014}.

The proposed method has several distinguishing characteristics, differentiating it from current users' reputation allocation procedures: (i) The method assigns users' reputation by considering the grouping behavior of users instead of based on the estimation of objects' true qualities. (ii) The method is with high performance in both accuracy and robustness, especially when dealing with small-degree spammers' attacks. (iii) The method is very efficient, as its time complexity \cite{Sipser2012} is $O(m^{2})$, which is significantly lower than most of previously proposed iterative methods. As further improvement, we could consider introducing this method to an iterative process \cite{Laureti2006,de2007,Zhou2011,Liao2014}, applying it to continuous rating systems \cite{Stanley2007,Treiblmaier2011}, and considering the effect of long-term evolution of online rating systems \cite{Zhao2014}.

\acknowledgments
This work was partially supported by the National Natural Science Foundation of China under Grant Nos. 61370150, 91324002, 61433014 and 11222543. J.G. acknowledges support from Tang Lixin Education Development Foundation by UESTC. T.Z. acknowledges the Program for New Century Excellent Talents in University under Grant No. NCET-11-0070, and Special Project of Sichuan Youth Science and Technology Innovation Research Team under Grant No. 2013TD0006.


\begin{thebibliography}{99}

\bibitem{Lu2012}
\Name{L{\"u} L., Medo M., Yeung C. H., Zhang Y.-C., Zhang Z.-K. \and Zhou T.}
\REVIEW{Phys. Rep.}{519}{2012}{1}.

\bibitem{Zeng2012a}
\Name{Zhang F., \and Zeng A.}
\REVIEW{EPL}{100}{2012}{58005}.

\bibitem{Zeng2014}
\Name{Zeng A., Vidmer A., Medo M. \and Zhang Y.-C.}
\REVIEW{EPL}{105}{2014}{58002}.

\bibitem{Guo2014}
\Name{Guo Q., Song W.-J. \and Liu J.-G.}
\REVIEW{EPL}{107}{2014}{18001}.

\bibitem{Jindal2007}
\Name{Jindal N. \and Liu B.} in
\Book{Proceedings of the 16th International Conference on World Wide Web}
 (ACM Press) 2007 pp. 1189--1190.

\bibitem{Muchnik2013}
\Name{Muchnik L., Aral S., \and Taylor S.-J.}
\REVIEW{Science}{341}{2013}{647}.

\bibitem{Yu2012}
\Name{Yu X., Liu Y., Huang X. \and An A.}
\REVIEW{IEEE Trans. on Knowl. and Data Eng.}{24}{2012}{720}.

\bibitem{Pan2013}
\Name{Pan W., Xiang E.-W., Liu N.-N. \and Yang Q.} in
\Book{Proceedings of the 24th AAAI Conference on Artificial Intelligence}
 (AAAI Press) 2013 pp. 39--55.

\bibitem{Wang2011}
\Name{Wang G., Xie S., Liu B. \and Yu P. S.} in
\Book{Proceedings of IEEE 11th International Conference on Data Mining (ICDM)} (IEEE Press) 2011 pp. 1242--1247.

\bibitem{Chung2013}
\Name{Chung C.-Y., Hsu P.-Y. \and Huang S.-H.}
\REVIEW{Decis. Support Syst.}{55}{2013}{314}.

\bibitem{Yang2012}
\Name{Yang Z., Zhang Z.-K. \and Zhou T.}
\REVIEW{EPL}{100}{2012}{68002}.

\bibitem{Huang2012}
\Name{Huang J., Cheng X.-Q., Shen H.-W., Zhou T., \and Jin X.} in
\Book{Proceedings of the 5th ACM International Conference on Web Search and Data Mining (WSDM)} (ACM Press) 2012 pp. 573--582.

\bibitem{Zhang2015}
\Name{Zhang Y.-L., Ni J., Guo Q., \and Liu J.-G.}
\REVIEW{Physica A}{417}{2015}{261}.

\bibitem{Zeng2012b}
\Name{Zeng A. \and Cimini G.}
\REVIEW{Phys. Rev. E}{85}{2012}{036101}.

\bibitem{Zhang2013}
\Name{Zhang Q.-M., Zeng A. \and Shang M.-S.}
\REVIEW{PLoS ONE}{8}{2013}{e62624}.

\bibitem{Zhao2014}
\Name{Zhao D.-D., Zeng A., Shang M.-S, \and Gao J.}
\REVIEW{Chin. Phys. Lett.}{30}{2013}{8901}.

\bibitem{Mukherjee2011}
\Name{Mukherjee A., Liu B., Wang J., Glance N., \and Jindal N.} in
\Book{Proceedings of the 20th International Conference on World Wide Web} (ACM Press) 2011 pp. 93--94.

\bibitem{Fei2013}
\Name{Fei G., Mukherjee A., Liu B., Hsu M., Castellanos M., \and Ghosh R.} in
\Book{Proceedings of the 7th International AAAI Conference on Weblogs and Social Media (ICWSM)} (AAAI Press) 2013 pp. 175--184.

\bibitem{Lin2014}
\Name{Lin Y., Zhu T., Wang X., Zhang J. \and Zhou A.} in
\Book{Proceedings of the Companion Publication of the 23rd International Conference on World Wide Web Companion} (WWW) 2014 pp. 341--342.

\bibitem{Masum2004}
\Name{Masum H. \and Zhang Y.-C.}
\REVIEW{First Monday}{9}{2004}{}.

\bibitem{Resnick2000}
\Name{Resnick P., Kuwabara K., Zeckhauser R. \and Friedman E.}
\REVIEW{Commun. ACM}{43}{2000}{45}.

\bibitem{Josang2007}
\Name{J{\o}sang A., Ismail R. \and Boyd C.}
\REVIEW{Decis. Support Syst.}{43}{2007}{618}.

\bibitem{Allahbakhsh2013}
\Name{Allahbakhsh M., Ignjatovic A., Motahari-Nezhad H. R. \and Benatallah B.}
\REVIEW{World Wide Web}{}{2013}{1}.

\bibitem{Ling2013}
\Name{Ling G., King I. \and Lyu M. R.} in
\Book{Proceedings of the 23rd International Joint Conference on Artificial Intelligence} (AAAI Press) 2013 pp. 2670--2676.

\bibitem{Laureti2006}
\Name{Laureti P., Moret L., Zhang Y.-C. \and Yu Y.-K.}
\REVIEW{EPL}{75}{2006}{1006}.

\bibitem{de2007}
\Name{de Kerchove C. \and  Van Dooren P.} arXiv:0711.3964 (2007)

\bibitem{Zhou2011}
\Name{Zhou Y.-B., Lei T. \and Zhou T.}
\REVIEW{EPL}{94}{2011}{48002}.

\bibitem{Liao2014}
\Name{Liao H., Zeng A., Xiao R., Ren Z.-M., Chen D.-B. \and Zhang Y.-C.}
\REVIEW{PLoS ONE}{9}{2014}{e97146}.

\bibitem{Tian2012}
\Name{Tian Y. \and Zhu J.} in
\Book{Proceedings of the 18th ACM SIGKDD International Conference on Knowledge Discovery and Data Mining} (ACM Press) 2012 pp. 226--234.

\bibitem{Raykar2010}
\Name{Raykar V. C., Yu S., Zhao L.-H., Valadez G. H., Florin C., Bogoni L. \and Moy L.}
\REVIEW{J. Mach. Learn. Res.}{11}{2010}{1297}.

\bibitem{Shi2009}
\Name{Shi X., Zhu J., Cai R., \and Zhang L.} in
\Book{Proceedings of the 15th ACM SIGKDD International Conference on Knowledge Discovery and Data Mining} (ACM Press) 2009 pp. 777--786.

\bibitem{Welinder2010}
\Name{Welinder P., Branson S., Belongie S. \and Perona P.} in
\Book{Advances in Neural Information Processing Systems (NIPS)} (MIT Press) 2010 pp. 2424--2432.

\bibitem{Reed2010}
\Name{Reed M. S., Evely A. C., Cundill G., Fazey I., Glass J., Laing A., Newig J., Parrish B., Prell C., Raymond C. \and Stringer L. C.}
\REVIEW{Ecology and Society}{15}{2010}{r1}.

\bibitem{Ross2010}
\Name{Ross J., Irani L., Silberman M., Zaldivar A. \and Tomlinson B.} in
\Book{CHI '10 Extended Abstracts on Human Factors in Computing Systems}
 (ACM Press) 2012 pp. 2863--2872.

\bibitem{Raafat2009}
\Name{Raafat R. M., Chater N., \and Frith C.}
\REVIEW{Trends Cogn. Sci.}{13}{2009}{420}.

\bibitem{Shang2010}
\Name{Shang M.-S., L{\"u} L., Zhang Y.-C. \and Zhou T.}
\REVIEW{EPL}{90}{2010}{48006}.

\bibitem{Zhou2007b}
\Name{Zhou T., Ren J., Medo M. \and Zhang Y.-C.}
\REVIEW{Phys. Rev. E}{76}{2007}{46115}.

\bibitem{Lin1989}
\Name{Lin L. I.}
\REVIEW{Biometrics}{45}{1989}{255}.

\bibitem{Jindal2008}
\Name{Jindal N. \and Liu B.} in
\Book{Proceedings of the 2008 International Conference on Web Search and Data Mining (WSDM)} (ACM Press) 2008 pp. 219--230.

\bibitem{Herlocker2004}
\Name{Herlocker J. L., Konstan J. A., Terveen L. G. \and Riedl J. T.}
\REVIEW{ACM Trans. Inf. Syst.}{22}{2004}{5}.

\bibitem{Hanley1982}
\Name{Hanley J. A., \and McNeil B. J.}
\REVIEW{Radiology}{143}{1982}{29}.

\bibitem{lu2011}
\Name{L{\"u}, L. \and Zhou T.}
\REVIEW{Physica A}{390}{2011}{1150}.

\bibitem{Sipser2012}
\Name{Sipser M.}
\Book{Introduction to the Theory of Computation} (Cengage Learning, Boston) 2012.

\bibitem{Stanley2007}
\Name{Stanley N. \and Jenkins S.} in
\Book{Challanges of a changing world. Proceedings of the 5th International Conference of the Association for Survey Computing} (Association for Survey Computing, Berkeley) 2007 pp. 81--92.

\bibitem{Treiblmaier2011}
\Name{Treiblmaier H. \and Filzmoser P.} in
\Book{Proceedings of the International Conference on Information Systems (ICIS)} (Association for Information Systems, Utrecht) 2011 Paper 1.

\end{thebibliography}

\expandafter\ifx\csname url\endcsname\relax\def\url#1{\texttt{#1}}\fi

\end{document}